\documentclass[11pt,a4paper]{article}

\usepackage{amsfonts,amsmath,amssymb,bm,cite}

\usepackage[active]{srcltx}

\begin{document}

\title{\textbf{Massive Majorana neutrinos in matter and a magnetic
field}}

\author{Maxim~Dvornikov$^{a,b}$\footnote{{\textbf{e-mail}}: maxdvo@izmiran.ru}
\\
$^a$\small{\emph{N.~V.~Pushkov Institute of Terrestrial Magnetism,
Ionosphere and Radiowave Propagation}} \\
\small{\emph{142190, Troitsk, Moscow Region, Russia}} \\
$^b$\small{\emph{Institute of Physics, University of S\~{a}o Paulo}} \\
\small{\emph{CP 66318, CEP 05315-970 S\~{a}o Paulo, SP, Brazil}}}
\date{}

\maketitle

\begin{abstract}
We summarize our recent results on the description of the
evolution of Majorana neutrinos in external fields. First we
discuss the purely quantum method which involves the secondly
quantized Weyl spinors. In frames of this formalism we find exact
solutions of the classical Hamilton equations for Weyl fields in
background matter, which then are used to canonically quantize the
system. The analog of the effective Hamiltonian for the
description of spin-flavor oscillations of Majorana neutrinos in
matter and a magnetic field is also obtained. Finally we discuss
another approach for the treatment of Majorana neutrinos in
external fields, which is based on the relativistic quantum
mechanics. The latter method does not require the quantization of
the neutrino wave function. Nevertheless we demonstrate that the
relativistic quantum mechanics approach also appropriately
describes spin-flavor oscillations of relativistic Majorana
neutrinos.
\end{abstract}

\section{Introduction}

It is known that the neutrino interaction with external fields can
significantly influence the evolution of
astrophysical~\cite{Raf96} and cosmological media~\cite{BerGoo04}.
For example, the interaction of supernova neutrinos with a strong
magnetic field of a protoneutron star (PNS) may cause a
macroscopic effect like the kick of a pulsar~\cite{KusSer96}. The
neutrino scattering off the matter of a rotating PNS may result in
a spin-down of a star~\cite{Mik77}. However the recent
estimates~\cite{DvoDib10} showed that the role of neutrinos in PNS
spin-down is significantly less important than that in the
possible explanation of great linear velocities of neutron stars.
The neutrino interaction with primordial plasma is essential for
the observed spectrum of the cosmic microwave
background~\cite{Han04} and for the generation of large scale
magnetic fields~\cite{SemSok04}.

It was experimentally confirmed that neutrinos are massive
particles and there is a mixing between different neutrino
generations~\cite{Abe08}. These neutrino properties result in the
appearance of neutrino flavor oscillations~\cite{GiuKim07p245}. It
is also known that the presence of background matter can cause the
resonance enhancement of neutrino
oscillations~\cite{GiuKim07p322}. Moreover the combined action of
a background matter and an external magnetic field leads to the
resonant transition like, $\nu_\alpha^{-{}} \leftrightarrow
\nu_\beta^{+{}}$ (see, e.g., Ref.~\cite{LimMar88}), where the
indexes $\pm{}$ denote different helicity states. Hence active
neutrinos of the flavor $\alpha$ can be converted into sterile
neutrinos of another flavor $\beta$.

Neutrinos can interact with an external electromagnetic field due
to the presence of anomalous magnetic moments~\cite{GiuStu09}.
Note that the structure of the magnetic moments is completely
different for Dirac and Majorana neutrinos (see, e.g.,
Ref.~\cite{FukYan03p461}). Despite the fact that nowadays there is
no universally recognized confirmation of the nature of
neutrinos~\cite{EllEng04}, in the present work we shall suppose
that neutrinos are Majorana particles. Note that in various
scenarios for the generation of elementary particles masses it is
predicted that neutrinos should acquire Majorana
masses~\cite{MohSmi06}. It should be also mentioned that numerous
experimental attempts are made to investigate whether neutrinos
are Dirac or Majorana particles~\cite{0nu2beta}.

In the present work we summarize our recent
results~\cite{Dvo11NOVA,Dvo11NPB} on the theory of Majorana
neutrinos oscillations in matter and electromagnetic fields.
First, in Sec.~\ref{NMV}, following the discussion of
Ref.~\cite{GiuKim07p180}, we review the general properties of the
neutrino mixing in vacuum. Then, in Sec.~\ref{ELECTROD}, we recall
the basic equation for a massive Majorana neutrino propagating in
background matter and electromagnetic field. In Sec.~\ref{CFT},
using the results of our recent works~\cite{Dvo11NPB,Dvo11} we
propose the classical field theory treatment of massive Weyl
fields propagating in a background matter and interacting with an
external electromagnetic field. Exact solutions of the wave
equation for Weyl fields in a background matter are found in
Sec.~\ref{ES}. In Sec.~\ref{QUANT} we use these solutions to
canonically quantize the Weyl fields. In Sec.~\ref{MATTEMF}, in
frames of our method we re-derive the effective Hamiltonian for
the description of spin-flavor oscillations of Majorana neutrinos
in matter and a magnetic field. Finally, in Sec.~\ref{RQM}, we
consider the alternative approach for the description of
spin-flavor oscillations of Majorana neutrinos based on the
relativistic quantum mechanics and compare it with the results of
Sec.~\ref{MATTEMF}. In Sec.~\ref{CONCL} we discuss our results.

\section{Neutrino mixing in vacuum\label{NMV}}

Experimentally it was found~\cite{Sch06} that the number of active
neutrinos is equal to three with the high level of accuracy.
Nevertheless in various theoretical and phenomenological scenarios
(see, e.g., Ref.~\cite{Moh04}) some additional sterile neutrinos
are discussed. Thus the general dynamics of the system of flavor
neutrinos can be formulated in terms of the $3 + N_s$ spinor
fields, where $N_s \geq 0$ is the number of sterile neutrinos. It
was experimentally confirmed~\cite{GolGroSun58} that active
neutrinos correspond to the left-handed chiral projections,
$\nu_\lambda^\mathrm{L}$, $\lambda = e, \mu, \tau$, whereas
sterile neutrinos should be expressed in terms of the right-handed
chiral projections, $\nu_\lambda^\mathrm{R}$, $\lambda = 1, \dots,
N_s$. Here $\nu_\mathrm{L,R} = (1 \mp \gamma^5)/2 \times \nu$.

The system of wave equations for the spinors
$\nu_\lambda^\mathrm{L,R}$ will involve both Dirac and Majorana
types of the mass terms. The simultaneous diagonalization of the
mass terms can be performed with help of the matrix
transformation~\cite{Kob80},
\begin{equation}\label{flmassrel}
  N_\lambda^\mathrm{L} =
  \sum_{a = 1}^{3+N_s} U_{\lambda a} \chi_a^\mathrm{L},
\end{equation}
where $N_\mathrm{L}^\mathrm{T} = (\nu_e^\mathrm{L},
\nu_\mu^\mathrm{L}, \nu_\tau^\mathrm{L}, [\nu_1^\mathrm{R}]^c,
\dots, [\nu_{N_s}^\mathrm{R}]^c)$ is the multiplet of the neutrino
fields, $(U_{\lambda a})$ is the unitary vacuum mixing matrix of
the size $(3+N_s) \times (3+N_s)$, and $\chi_a$ are the neutrino
mass eigenstates. The charge conjugation is defined in the
standard manner: $\nu^c = \mathrm{i} \gamma^2 \nu^{*{}}$.

The fields $\chi_a$ are neither Dirac nor Majorana. However we may
always construct Majorana fields of the the chiral projections of
$\chi_a$ as
\begin{equation}\label{Majfieldsdef}
  \psi_a =
  \left[
    \varkappa^{*{}} \chi_a^\mathrm{L} +
    \tilde{\varkappa}^{*{}} (\chi_a^\mathrm{L})^c
  \right],
\end{equation}
where $\varkappa$ and $\tilde{\varkappa}$ are the phase factors
having unit absolute values. Now we can see that the fields
$\psi_a$ satisfy the Majorana condition in an extended sense,
\begin{equation}\label{Majcond}
  \psi_a^c = \varkappa_c \psi_a,
\end{equation}
where $\varkappa_c = \varkappa \tilde{\varkappa}$. In the
following we shall suppose that $\varkappa_c = 1$.

\section{Electrodynamics of Majorana neutrinos in a background matter\label{ELECTROD}}

The evolution of Majorana fields $\psi_a$, defined in
Eq.~\eqref{Majfieldsdef}, in a background matter under the
influence of an external electromagnetic field is governed by the
following wave equation:
\begin{equation}\label{wepsi}
  (\mathrm{i} \gamma^\mu \partial_\mu - m_a) \psi_a -
  \frac{\mu_{ab}}{2} \sigma_{\mu\nu} F^{\mu\nu} \psi_b +
  g_{ab}^\mu \gamma_\mu \gamma^5 \psi_b = 0,
\end{equation}
where $m_a$ are the masses of the particles and $\sigma_{\mu\nu} =
(\mathrm{i}/2)[\gamma_\mu; \gamma_\nu]_{-{}}$. Note that we will
formulate the dynamics of the system~\eqref{wepsi} in the mass
eigenstates basis rather than in the flavor basis, as it is
usually done when neutrino oscillations are considered, since, as
we demonstrated in Sec.~\ref{NMV}, only in the mass eigenstates
basis one can distinguish between Dirac and Majorana masses.

The matrix of the neutrino interaction with a background matter,
$(g_{ab}^\mu)$, introduced in Eq.~\eqref{wepsi} in the mass
eigenstates basis, is related to the analogous matrix in the
flavor basis as follows:
\begin{equation}\label{mattintmatrrel}
  g_{ab}^\mu = \sum_{\lambda, \lambda' = e, \mu, \tau}
  U^\dag_{a\lambda} f_{\lambda\lambda'}^\mu U_{\lambda'b},
\end{equation}
where the mixing matrix $(U_{\lambda a})$ is defined in
Eq.~\eqref{flmassrel}. In general case the matrix $(g_{ab}^\mu)$
is hermitian. However we shall discuss the situation when the CP
invariance is conserved. In this case the matrix $(g_{ab}^\mu)$ is
symmetric. Despite a current attempt to detect CP violating terms
in the neutrino sector~\cite{Abe11}, no definite results have been
obtained yet.

Note that in Eq.~\eqref{mattintmatrrel} we do not exclude the
possibility of the existence of nonstandard neutrino interactions,
which correspond to the nondiagonal elements of the matrix
$(f_{\lambda\lambda'}^\mu)$ (see, e.g., Ref.~\cite{Dvo11NOVA}). In
the situation when only the standard model interactions with
matter are present, the matrix $(f_{\lambda\lambda'}^\mu)$ is
diagonal: $f_{\lambda\lambda'}^\mu = \delta_{\lambda\lambda'}
f_\lambda^\mu$, where $f_\lambda^\mu$ are the effective potentials
of interactions with background fermions. The zero components of
these potentials, $f_\lambda^0$, are proportional to the effective
matter density of non-moving and unpolarized background fermions,
whereas the vector components, $\mathbf{f}_\lambda$, are the
linear combinations of the averaged matter velocity and the
polarization. The explicit form of the effective potentials and
the details of the statistical averaging can be found in
Ref.~\cite{DvoStu02}.

Note that the vector term in the neutrino matter interaction $\sim
g_{ab}^\mu \gamma_\mu \psi_b$ is omitted in Eq.~\eqref{wepsi}
since it is washed out for Majorana neutrinos. The contribution of
the axial-vector interaction with matter $\sim g_{ab}^\mu
\gamma_\mu \gamma^5 \psi_b$ to the wave equation~\eqref{wepsi} is
twice the analogous contribution for Dirac particles since both
neutrinos and antineutrinos equally interact with a background
matter (see, e.g., Ref.~\cite{GriStuTer05}).

Neutrinos can interact with the external electromagnetic field
$F_{\mu\nu} = (\mathbf{E},\mathbf{B})$ due to the presence of the
anomalous magnetic moments $(\mu_{ab})$. It is know (see, e.g.,
Ref.~\cite{PasSegSemVal00}) that the matrix $(\mu_{ab})$ should be
hermitian and pure imaginary, i.e. $\mu_{ab} = - \mu_{ba}$ and
$\mu_{ab}^{*{}} = - \mu_{ab}$. Unlike the interaction with matter,
which is generically defined for flavor neutrinos, the computation
of the neutrino magnetic moments has to be done in the mass
eigenstates basis. The detailed example of such a calculation is
given in Ref.~\cite{DvoStu04}. We shall discuss the situation when
no admixture of sterile neutrinos is in the mass eigenstates
$\psi_a$, i.e. when the eigenstates $\nu_\lambda^\mathrm{R}$ are
absent. In this case the neutrino electric dipole moments are
equal to zero~\cite{PasSegSemVal00}.

In Ref.~\cite{FukYan03p289} is was rigorously proved that four
component Majorana spinors are equivalent to two component Weyl
spinors. Thus we may re-express the wave function $\psi_a$ in
Eq.~\eqref{mattintmatrrel} in two ways,
\begin{equation}\label{psietaxi}
  \psi_a^{(\eta)} =
  \begin{pmatrix}
    \mathrm{i} \sigma_2 \eta_a^{*{}} \\
    \eta_a \
  \end{pmatrix},
  \quad
  \text{or}
  \quad
  \psi_a^{(\xi)} =
  \begin{pmatrix}
    \xi_a  \\
    - \mathrm{i} \sigma_2 \xi_a^{*{}} \
  \end{pmatrix}.
\end{equation}
Both representations in Eq.~\eqref{psietaxi} satisfy the Majorana
condition~\eqref{Majcond}. Using Eq.~\eqref{psietaxi} we can
rewrite Eq.~\eqref{wepsi} as
\begin{equation}\label{weleft}
  \dot{\eta}_a - (\bm{\sigma}\nabla)\eta_a +
  m_a \sigma_2 \eta_a^{*{}} -
  \mu_{ab} \bm{\sigma}
  (\mathbf{B} - \mathrm{i}\mathbf{E}) \sigma_2 \eta_b^{*{}} +
  \mathrm{i}
  (g^0_{ab}+\bm{\sigma}\mathbf{g}_{ab}) \eta_b = 0,
\end{equation}
or
\begin{equation}\label{weright}
  \dot{\xi}_a  + (\bm{\sigma}\nabla)\xi_a -
  m_a \sigma_2 \xi_a^{*{}} +
  \mu_{ab} \bm{\sigma}
  (\mathbf{B} + \mathrm{i}\mathbf{E}) \sigma_2 \xi_b^{*{}} -
  \mathrm{i}
  (g^0_{ab}-\bm{\sigma}^{*{}}\mathbf{g}_{ab}) \xi_b = 0.
\end{equation}
In the following we shall postulate these equations. Note that the
analog of Eq.~\eqref{weleft} was previously derived in
Ref.~\cite{DvoMaa09}.

\section{Classical field theory\label{CFT}}

In Ref.~\cite{Dvo11} we demonstrated that the classical dynamics
of a massive Weyl field in vacuum should be described only in
frames of the Hamilton formalism. Generalizing the results of
Ref.~\cite{Dvo11} to include the interactions with a background
matter and an electromagnetic field we arrive to the following
Hamiltonian (see also Ref.~\cite{Dvo11NPB}):
\begin{align}\label{Hamclass}
  H = & \int \mathrm{d}^3\mathbf{r}
  \Big[
  \sum_a
  \left\{
    \pi_a^\mathrm{T} (\bm{\sigma}\nabla) \eta_a -
    (\eta_a^{*{}})^\mathrm{T} (\bm{\sigma}\nabla) \pi_a^{*{}}
    +
    m_a
    \left[
      (\eta_a^{*{}})^\mathrm{T} \sigma_2 \pi_a +
      (\pi_a^{*{}})^\mathrm{T} \sigma_2 \eta_a
    \right]
  \right\}
  \notag
  \\
  & +
  \sum_{ab}
  \big\{
    \mu_{ab}
    \left[
      \pi_a^\mathrm{T} \bm{\sigma}
      (\mathbf{B} - \mathrm{i}\mathbf{E}) \sigma_2 \eta_b^{*{}} +
      \eta_a^\mathrm{T} \sigma_2 \bm{\sigma}
      (\mathbf{B} + \mathrm{i}\mathbf{E}) \pi_b^{*{}}
    \right]
    \notag
    \\
    & -
    \mathrm{i}
    \left[
      \pi_a^\mathrm{T}
      (g^0_{ab}+\bm{\sigma}\mathbf{g}_{ab}) \eta_b -
      (\eta_a^{*{}})^\mathrm{T}
      (g^0_{ab}+\bm{\sigma}\mathbf{g}_{ab}) \pi_b^{*{}}
    \right]
  \big\}
  \Big],
\end{align}
where $\pi_a$ are the canonical momenta conjugate to the
``coordinates" $\eta_a$. Using the aforementioned properties of
the matrices $(\mu_{ab})$ and $(g_{ab}^\mu)$ we find that the
functional~\eqref{Hamclass} is real as it should be for a
classical Hamiltonian.

Applying the field theory version of the canonical equations to
the Hamiltonian $H$,
\begin{align}
  \label{etaclass}
  \dot{\eta}_a  = & \frac{\delta H}{\delta \pi_a} =
  (\bm{\sigma}\nabla)\eta_a - m_a \sigma_2 \eta_a^{*{}} +
  \mu_{ab} \bm{\sigma}
  (\mathbf{B} - \mathrm{i}\mathbf{E}) \sigma_2 \eta_b^{*{}} -
  \mathrm{i} (g^0_{ab}+\bm{\sigma}\mathbf{g}_{ab}) \eta_b,
  \\
  \label{piclass}
  \dot{\pi}_a  = & - \frac{\delta H}{\delta \eta_a} =
  (\bm{\sigma}^{*{}}\nabla)\pi_a + m_a \sigma_2 \pi_a^{*{}} -
  \mu_{ab} \sigma_2 \bm{\sigma}
  (\mathbf{B} + \mathrm{i}\mathbf{E}) \pi_b^{*{}} +
  \mathrm{i} (g^0_{ab}+\bm{\sigma}^{*{}}\mathbf{g}_{ab}) \pi_b,
\end{align}
one can see that in Eq.~\eqref{etaclass} we reproduce
Eq.~\eqref{weleft} for Weyl particles, which correspond to
left-handed neutrinos, interacting with matter and an
electromagnetic field. If we introduce the new variable $\xi_a =
\mathrm{i} \sigma_2 \pi_a$, we can show that Eq.~\eqref{piclass}
is equivalent to Eq.~\eqref{weright} for right-handed neutrinos.

\section{Exact solution of the wave equation\label{ES}}

In the following we shall suppose that the background matter in
average is at rest and unpolarized, i.e. $\mathbf{g}_{ab} = 0$.
This approximation is valid in almost all realistic cases (see the
detailed discussion in Ref.~\cite{Dvo11NPB}). Let us decompose the
Hamiltonian~\eqref{Hamclass} into two terms $H = H_0 +
H_\mathrm{int}$. The former term, $H_0$, contains the vacuum
Hamiltonian as well as the matter term diagonal in neutrino types,
\begin{align}\label{Ham0}
  H_0 = & \int \mathrm{d}^3\mathbf{r}
  \sum_a
  \big\{
    \pi_a^\mathrm{T} [(\bm{\sigma}\nabla) - \mathrm{i} g^0_{aa}] \eta_a -
    (\eta_a^{*{}})^\mathrm{T} [(\bm{\sigma}\nabla) - \mathrm{i} g^0_{aa}] \pi_a^{*{}}
    \notag
    \\
    & +
    m_a
    \left[
      (\eta_a^{*{}})^\mathrm{T} \sigma_2 \pi_a +
      (\pi_a^{*{}})^\mathrm{T} \sigma_2 \eta_a
    \right]
  \big\}.
\end{align}
The latter term in this decomposition,
\begin{equation}\label{Hamint}
  H_\mathrm{int} = \int \mathrm{d}^3\mathbf{r}
  \sum_{a \neq b}
  \big\{
    \mu_{ab}
    \left[
      \pi_a^\mathrm{T} \bm{\sigma}
      (\mathbf{B} - \mathrm{i}\mathbf{E}) \sigma_2 \eta_b^{*{}} +
      \eta_a^\mathrm{T} \sigma_2 \bm{\sigma}
      (\mathbf{B} + \mathrm{i}\mathbf{E}) \pi_b^{*{}}
    \right] -
    \mathrm{i} g^0_{ab}
    \left[
      \pi_a^\mathrm{T} \eta_b -
      (\eta_a^{*{}})^\mathrm{T} \pi_b^{*{}}
    \right]
  \big\},
\end{equation}
has the nondiagonal matter interaction and the interaction with an
electromagnetic field which is nondiagonal by definition.

Analogously to Eqs.~\eqref{etaclass} and~\eqref{piclass} we define
the reduced Hamilton equations which contain only the Hamiltonian
$H_0$: $\dot{\eta}_a^{(0)} = \delta H_0/\delta \pi_a^{(0)}$ and
$\dot{\pi}_a^{(0)} = - \delta H_0/\delta \eta_a^{(0)}$. Using the
results of Refs.~\cite{Dvo11NPB,DvoMaa09} we can find the
solutions of these equations in the form,
\begin{align}\label{etaxisol}
  \eta_a^{(0)}(\mathbf{r},t) = & \frac{1}{2}
  \int \frac{\mathrm{d}^3\mathbf{p}}{(2\pi)^{3/2}}
  \bigg\{
    \left[
      a_a^{-{}} w_{-{}} e^{-\mathrm{i}E_a^{-{}}t} -
      \frac{m_a}{E_a^{+{}}+|\mathbf{p}|-g^0_{aa}}
      a_a^{+{}} w_{+{}} e^{-\mathrm{i}E_a^{+{}}t}
    \right]
    e^{\mathrm{i}\mathbf{p}\mathbf{r}}
    \notag
    \\
    & +
    \left[
      (a_a^{+{}})^{*{}} w_{-{}} e^{\mathrm{i}E_a^{+{}}t} +
      \frac{m_a}{E_a^{-{}}+|\mathbf{p}|+g^0_{aa}}
      (a_a^{-{}})^{*{}} w_{+{}} e^{\mathrm{i}E_a^{-{}}t}
    \right]
    e^{-\mathrm{i}\mathbf{p}\mathbf{r}}
  \bigg\},
  \notag
  \displaybreak[2]
  \\
  \xi_a^{(0)}(\mathbf{r},t) = & \frac{\mathrm{i}}{2}
  \int \frac{\mathrm{d}^3\mathbf{p}}{(2\pi)^{3/2}}
  \bigg\{
    \left[
      b_a^{+{}} w_{+{}} e^{-\mathrm{i}E_a^{+{}}t} +
      \frac{m_a}{E_a^{-{}}+|\mathbf{p}|+g^0_{aa}}
      b_a^{-{}} w_{-{}} e^{-\mathrm{i}E_a^{-{}}t}
    \right]
    e^{\mathrm{i}\mathbf{p}\mathbf{r}}
    \notag
    \\
    & +
    \left[
      (b_a^{-{}})^{*{}} w_{+{}} e^{\mathrm{i}E_a^{-{}}t} -
      \frac{m_a}{E_a^{+{}}+|\mathbf{p}|-g^0_{aa}}
      (b_a^{+{}})^{*{}} w_{-{}} e^{\mathrm{i}E_a^{+{}}t}
    \right]
    e^{-\mathrm{i}\mathbf{p}\mathbf{r}}
  \bigg\},
\end{align}
where we introduce the new variable $\xi_a^{(0)} = \mathrm{i}
\sigma_2 \pi_a^{(0)}$, $w_{\pm{}}$ are the helicity amplitudes
defined in Ref.~\cite{BerLifPit82}, and
\begin{equation}\label{enWeyl}
  E_a^{(\zeta)} =
  \sqrt{m_a^2 + (|\mathbf{p}| - \zeta g^0_{aa})^2}.
\end{equation}
is the energy of a Weyl field~\cite{GriStuTer05}, $\zeta = \pm 1$
is the particle helicity. To derive Eqs.~\eqref{etaxisol}
and~\eqref{enWeyl} we suppose that the external field $g^0_{aa}$
is spatially constant.

\section{Quantization\label{QUANT}}

Now we can carry out the canonical quantization of the Weyl fields
$\eta_a$ and $\pi_a$. For this purpose we suggest that the
expansion coefficients $a_a^{\pm{}}(\mathbf{p})$ and
$b_a^{\pm{}}(\mathbf{p})$ in Eq.~\eqref{etaxisol} are operators.
Note that the operators in the expansion of $\eta_a$ and $\xi_a$
are independent since in Sec.~\ref{CFT} we showed that these
fields evolve independently.

Using the following relation:
\begin{equation}\label{Majcondquant}
  a_a^{\pm{}}(\mathbf{p})(E_a^{\pm{}} + |\mathbf{p}| \mp g^0_{aa}) =
  b_a^{\pm{}}(\mathbf{p})(|\mathbf{p}| \mp g^0_{aa}),
\end{equation}
and suggesting that the operators $a_a^{\pm{}}(\mathbf{p})$ obey
the anti-commutation properties,
\begin{equation}\label{anticomm}
  \{
    a_a^{\pm{}}(\mathbf{k}); [a_b^{\pm{}}(\mathbf{p})]^{*{}}
  \}_{+{}} =
  \delta_{ab} \delta^3(\mathbf{k} - \mathbf{p}),
\end{equation}
with all the rest of the anticommutators being equal to zero, we
can cast Eq.~\eqref{Ham0} into the form,
\begin{equation}\label{totenquant}
  H_0 = \int \mathrm{d}^3\mathbf{p}
  \sum_a
  [E_a^{-{}} (a_a^{-{}})^{*{}} a_a^{-{}} +
  E_a^{+{}} (a_a^{+{}})^{*{}} a_a^{+{}}] + \text{divergent terms},
\end{equation}
which shows that the total energy of a massive Weyl field is a sum
of energies corresponding to elementary oscillators of positive
and negative helicities. Note that the details of the derivation
of Eq.~\eqref{totenquant} can be found in Ref.~\cite{Dvo11NPB}

Using the results Ref.~\cite{Dvo11} we can also quantize the total
momentum of a Weyl field defined as
\begin{equation}\label{momdef}
  \mathbf{P}_0 = \int \mathrm{d}^3\mathbf{r}
  \sum_a
  \left[
    \left( \eta_a^{(0)*{}} \right)^\mathrm{T} \nabla \pi_a^{(0)*{}} -
    \left( \pi_a^{(0)} \right)^\mathrm{T} \nabla \eta_a^{(0)}
  \right].
\end{equation}
With help of Eqs.~\eqref{etaxisol}, \eqref{Majcondquant},
and~\eqref{anticomm} we rewrite Eq.~\eqref{momdef} in the
following form:
\begin{equation}\label{totmomquant}
  \mathbf{P}_0 = \int \mathrm{d}^3\mathbf{p}
  \sum_a
  \mathbf{p}
  [(a_a^{-{}})^{*{}} a_a^{-{}} +
  (a_a^{+{}})^{*{}} a_a^{+{}}] + \text{divergent terms},
\end{equation}
which has the similar structure as Eq.~\eqref{totenquant}.

It is interesting to mention that a massive Weyl field in vacuum
can be quantized in the two independent ways (see
Ref.~\cite{Dvo11}) because of the degeneracy of the neutrino
energy levels: $E_a^{-{}} = E_a^{+{}} = \sqrt{m_a^2 +
|\mathbf{p}|^2}$. On the contrary, in matter only one of the
possibilities for the quantization gives the correct result for
the total energy~\eqref{totenquant} since the energy levels are no
longer degenerate, cf. Eq.~\eqref{enWeyl}.

\section{Nondiagonal interaction with matter and a magnetic field\label{MATTEMF}}

To quantize the Hamiltonian $H_\mathrm{int}$ we shall use the
forward scattering approximation. It means that one has to account
for only the terms conserving the number of
particles~\cite{Raf96p318}. Using  Eqs.~\eqref{etaxisol},
\eqref{Majcondquant}, and~\eqref{anticomm} we rewrite
Eq.~\eqref{Hamint} in the form,
\begin{align}\label{Hamintquant}
  H_\mathrm{int} = & \int \mathrm{d}^3\mathbf{p}
  \sum_{a \neq b}
  [M_{ab}^{-{}} (a_a^{-{}})^{*{}} a_b^{-{}}
  e^{\mathrm{i}\delta_{ab}^{-}t} +
  M_{ab}^{+{}} (a_a^{+{}})^{*{}} a_b^{+{}}
  e^{\mathrm{i}\delta_{ab}^{+}t}
  \notag
  \\
  & +
  F_{ab}^{-{}} (a_a^{-{}})^{*{}} a_b^{+{}}
  e^{\mathrm{i}\sigma_{ab}^{-}t} +
  F_{ab}^{+{}} (a_a^{+{}})^{*{}} a_b^{-{}}
  e^{\mathrm{i}\sigma_{ab}^{+}t}],
\end{align}
where $\delta_{ab}^{\pm{}} = E_a^{\pm{}} - E_b^{\pm{}}$ and
$\sigma_{ab}^{\pm{}} = E_a^{\pm{}} - E_b^{\mp{}}$.

The general form of the coefficients $M_{ab}^{\pm{}}$ and
$F_{ab}^{\pm{}}$ is quite cumbersome. Therefore we describe the
dynamics of the system in the ultrarelativistic approximation,
$|\mathbf{k}| \gg \max(m_{a}, g^0_{aa})$, where $\mathbf{k}$ is
the initial momentum of neutrinos. Moreover we discuss the
simplest case of the two neutrino eigenstates, $a=1,2$, and
suppose that $\mathbf{E} = 0$, since it is difficult to create a
large scale electric field. In this limit we get for the
coefficients $ M_{ab}^{\pm{}} \approx \mp g_{ab}$ and
$F_{ab}^{\pm{}} \approx - \mu_{ab} |\mathbf{B}| \sin
\vartheta_\mathbf{kB}$, where $\vartheta_\mathbf{kB}$ is the angle
between the vectors $\mathbf{k}$ and $\mathbf{B}$. The explicit
form of these coefficients for the arbitrary neutrino momentum can
be found in Ref.~\cite{Dvo11NPB}.

Now we define the neutrino density matrix as
\begin{equation}\label{rhodef}
  \delta^3(\mathbf{p}-\mathbf{k}) \rho_{AB}(\mathbf{k}) =
  \langle a^{*{}}_B(\mathbf{p}) a_A(\mathbf{k}) \rangle,
\end{equation}
where $A = (\zeta, a)$ is a composite index and $\langle \dots
\rangle$ is the statistical averaging over the neutrino ensemble.
Applying the quantum Liouville equation for the description of the
density matrix evolution,
\begin{equation}\label{qLiouv}
  \mathrm{i} \dot{\rho} = [\rho, H_\mathrm{int}],
\end{equation}
we can rewrite it as $\mathrm{i} \dot{\rho} = [\mathcal{H},
\rho]$, where
\begin{equation}\label{Hinteff}
  \mathcal{H} =
  \begin{pmatrix}
    (M_{ab}^{-{}} e^{\mathrm{i}\delta_{ab}^{-}t}) &
    (F_{ab}^{-{}} e^{\mathrm{i}\sigma_{ab}^{-}t}) \\
    (F_{ab}^{+{}} e^{\mathrm{i}\sigma_{ab}^{+}t}) &
    (M_{ab}^{+{}} e^{\mathrm{i}\delta_{ab}^{+}t}) \
  \end{pmatrix},
\end{equation}
is the effective quantum mechanical Hamiltonian.

To study the evolution of our system we use Eq.~\eqref{etaxisol},
where the wave functions already contain time dependent
exponential factors. The energies in Eq.~\eqref{etaxisol}
correspond to the total diagonal Hamiltonian~\eqref{Ham0}, cf.
Eqs.~\eqref{enWeyl} and~\eqref{totenquant}, which contains both
the mass term and the diagonal interaction with a background
matter rather than only a kinetic term as in Ref.~\cite{SigRaf93}.
Thus our treatment is similar to the Dirac picture of the quantum
theory. That is why in Eq.~\eqref{qLiouv} it is sufficient to
commute the density matrix only with $H_\mathrm{int}$ rather than
with the total Hamiltonian $H = H_0 + H_\mathrm{int}$.

To eliminate the time dependence in the effective
Hamiltonian~\eqref{Hinteff} we make the transformation of the
density matrix~\cite{DvoMaa09},
\begin{equation}\label{matrtransf}
  \rho_\mathrm{qm} =
  \mathcal{U} \rho \mathcal{U}^\dag,
  \quad
  \mathcal{U} = \mathrm{diag}\{e^{-\mathrm{i}(\Phi+g^0_{11})t},
  e^{\mathrm{i}(\Phi-g^0_{22})t},
  e^{-\mathrm{i}(\Phi-g^0_{11})t},
  e^{\mathrm{i}(\Phi+g^0_{22})}t\},
\end{equation}
where $\Phi = \delta m^2 / 4|\mathbf{k}|$ is the phase of vacuum
oscillations and $\delta m^2 = m_1^2 -m_2^2$ is the mass squared
difference.

The evolution of the transformed density matrix can be represented
as $\mathrm{i} \dot{\rho}_\mathrm{qm} = [\mathcal{H}_\mathrm{qm},
\rho_\mathrm{qm}]$, where the new effective Hamiltonian has the
form,
\begin{align}\label{Heffqm}
  \mathcal{H}_\mathrm{qm} = &
  \mathcal{U} \mathcal{H} \mathcal{U}^\dag +
  \mathrm{i} \dot{\mathcal{U}}\mathcal{U}^\dag
  \notag
  \\
  & =
  \begin{pmatrix}
    \Phi + g^0_{11} & g^0_{12} &
    0 & -\mu_{12} |\mathbf{B}| \sin \vartheta_\mathbf{kB} \\
    g^0_{21} & - \Phi + g^0_{22} &
    -\mu_{21} |\mathbf{B}| \sin \vartheta_\mathbf{kB} & 0 \\
    0 & -\mu_{12} |\mathbf{B}| \sin \vartheta_\mathbf{kB} &
    \Phi - g^0_{11} & - g^0_{12} \\
    -\mu_{21} |\mathbf{B}| \sin \vartheta_\mathbf{kB} & 0 &
    - g^0_{21} & - \Phi - g^0_{22} \
  \end{pmatrix}.
\end{align}
Recalling the properties of the magnetic moments matrix for
Majorana neutrinos, $\mu_{12} = \mathrm{i}\mu$ and $\mu_{21} = -
\mathrm{i}\mu$, with $\mu$ being a real number, we can see that
Eq.~\eqref{Heffqm} reproduces the well known quantum mechanical
Hamiltonian for spin-flavor oscillations of Majorana neutrinos in
matter and a magnetic field~\cite{LimMar88}.

\section{Relativistic quantum mechanics\label{RQM}}

In Sec.~\ref{MATTEMF} we showed that the dynamics of neutrino
spin-flavor oscillations in matter and a magnetic field can be
described in frames of the purely quantum approach based on the
quantum Liouville equation~\eqref{qLiouv} for the neutrino density
matrix~\eqref{rhodef}. There is, however, an alternative formalism
for the treatment of neutrino oscillations which involves the
relativistic quantum mechanics method~\cite{Dvo11NOVA,DvoMaa09}.
This approach is based on the exact solutions of wave equations
for massive neutrinos in an external field, which were obtained in
Sec.~\ref{ES}.

The main idea of the relativistic quantum mechanics method is the
following: for the given initial wave functions of flavor
neutrinos one should find the wave functions at subsequent moments
of time. Note that this approach is similar to the relativistic
wave packets description of neutrinos
oscillations~\cite{GiuKim07p299}. However we study the propagation
of neutrino wave packets which exactly account for external
fields. If one defines some known initial distribution of active
neutrinos, then we can calculate the transition and survival
probabilities for other neutrino flavors. Note that using
Eq.~\eqref{flmassrel} we can always express the initial condition
for flavor neutrinos via that for the neutrino mass eigenstates,
i.e. we shall suppose that $\eta_a(\mathbf{r}, t=0) =
\eta_a^{(0)}(\mathbf{r})$ are the given functions.

In our analysis we shall assume that the expansion coefficients in
the decomposition of the wave functions~\eqref{etaxisol} are
$c$-numbers rather than operators as it was supposed in
Sec.~\ref{QUANT}. Thus, since we do not quantize the Weyl fields,
we no longer need the additional degree of freedom represented by
the spinor $\xi$. Moreover we may change the normalization of the
spinors at our convenience.

We will look for a general solution of Eq.~\eqref{weleft} in the
following form (see Eq.~\eqref{etaxisol} and
Ref.~\cite{DvoMaa09}):
\begin{align}\label{etasolrqm}
  \eta_a(\mathbf{r},t) = &
  \int \frac{\mathrm{d}^3\mathbf{p}}{(2\pi)^{3/2}}
  \bigg\{
    \left(
      a_a^{-{}}(\mathbf{p},t) w_{-{}} e^{-\mathrm{i}E_a^{-{}}t} -
      \frac{m_a}{E_a^{+{}}+|\mathbf{p}|-g^0_{aa}}
      a_a^{+{}}(\mathbf{p},t) w_{+{}} e^{-\mathrm{i}E_a^{+{}}t}
    \right)
    e^{\mathrm{i}\mathbf{p}\mathbf{r}}
    \notag
    \\
    & +
    \left(
      [a_a^{+{}}(\mathbf{p},t)]^{*{}} w_{-{}} e^{\mathrm{i}E_a^{+{}}t} +
      \frac{m_a}{E_a^{-{}}+|\mathbf{p}|+g^0_{aa}}
      [a_a^{-{}}(\mathbf{p},t)]^{*{}} w_{+{}} e^{\mathrm{i}E_a^{-{}}t}
    \right)
    e^{-\mathrm{i}\mathbf{p}\mathbf{r}}
  \bigg\},
\end{align}
where the expansion coefficients $a_a^{\pm{}}(\mathbf{p},t)$ are
the functions of time to account for the nondiagonal terms in
Eq.~\eqref{weleft}. The energy levels $E_a^{\pm{}}$ and the
helicity amplitudes $w_{\pm{}}$ are defined in Sec.~\ref{ES}.

We choose the plane wave initial wave functions of massive
neutrinos, $\eta_a^{(0)}(\mathbf{r}) = A_a
e^{\mathrm{i}\mathbf{k}\mathbf{r}}$, where $A_a$ is the amplitude
of the initial neutrino field distribution. If we are not
interested in the next-to-leading effects in neutrino
oscillations, we may discuss the ultrarelativistic approximation
from the very beginning: $|\mathbf{k}| \gg \max(m_{a}, g^0_{aa})$.
As in Sec.~\ref{MATTEMF} we suggest that $\mathbf{E} = 0$ as well
as adopt the following geometry: $\mathbf{B} = B \mathbf{e}_z$ and
$\mathbf{k} = |\mathbf{k}| \times (\sin \vartheta_\mathbf{kB}, 0,
\cos \vartheta_\mathbf{kB})$.

Inserting the \textit{ansatz}~\eqref{etasolrqm} in
Eq.~\eqref{weleft} and after the straightforward calculations we
get the system of the ordinary differential equations for
$a_a^{\pm{}}(\mathbf{p},t)$. It is convenient to rewrite it in a
conventional form as the Schr\"{o}dinger-like equation,
$\mathrm{i} \dot{\Psi} = \mathcal{H} \Psi$, introducing the
quantum mechanical ``wave function" of neutrinos, $\Psi^\mathrm{T}
= (a_1^{-{}}, a_2^{-{}}, \dots, a_1^{+{}}, a_2^{+{}}, \dots)$, and
the effective Hamiltonian $\mathcal{H}$, which has the same form
as in Eq.~\eqref{Hinteff}. As in Sec.~\ref{MATTEMF}, here we shall
discuss the situation of two neutrino eigenstates, $a = 1, 2$.
Then we make the transformation, $\Psi_\mathrm{qm} = \mathcal{U}
\Psi$, where the matrix $\mathcal{U}$ is defined in
Eq.~\eqref{matrtransf}. The evolution of the transformed quantum
mechanical ``wave function" $\Psi_\mathrm{qm}$ is governed by the
effective Hamiltonian $\mathcal{H}_\mathrm{qm}$ which coincides
with that given in Eq.~\eqref{Heffqm}. Therefore we obtain that
the relativistic quantum mechanics method is consistent with the
standard quantum mechanical description of neutrino spin-flavor
oscillations in matter and a magnetic field.

\section{Conclusion\label{CONCL}}

In the present work we have summarized our recent results on the
theory of spin-flavor oscillations of Majorana neutrinos in matter
and a magnetic field. After the brief introduction to the theory
of neutrino mixing in vacuum in Sec.~\ref{NMV} and the discussion
of the Majorana neutrinos electrodynamics in Sec.~\ref{ELECTROD},
we have mainly considered two approaches: the purely quantum
method, Secs.~\ref{CFT}-\ref{MATTEMF}, and the relativistic
quantum mechanics approach, Sec.~\ref{RQM}.

The former method is based on the canonical quantization of Weyl
fields in presence of a background matter. The procedure of the
canonical quantization requires the construction of the classical
field theory Hamiltonian~\cite{Wei96}, in which all the dynamical
observables should then be replaced with operators. Previously it
was claimed~\cite{SchVal81} that massive Weyl fields are
essentially quantum objects to be expressed via anticommuting
operators. Nevertheless, in Sec.~\ref{CFT}, we have constructed
the classical field theory of massive Weyl fields in a background
matter and an electromagnetic field, which is based on the
Hamilton formalism. Note that one does not doubt that generically
our world is quantum. However there are numerous processes which
may be also described within the classical physics (see
Ref.~\cite{Gui03} for many interesting examples).

In Sec.~\ref{ES} we have found the exact solution of the wave
equation for a Weyl field in presence of a background matter. This
solution has been used in Sec.~\ref{QUANT} to canonically quantize
the system. Then, in Sec.~\ref{MATTEMF}, we have completed the
quantization of the system by taking into account the nondiagonal
interaction with matter and a magnetic field using the density
matrix formalism. For ultrarelativistic particles we have
re-derived the effective Hamiltonian for the description of
spin-flavor oscillations of Majorana neutrinos in matter and a
magnetic field.

Finally, in Sec.~\ref{RQM}, we have considered the alternative
approach for the description of the evolution of Majorana
neutrinos which is based on the relativistic quantum mechanics. In
frames of this method we do not quantize the neutrino wave
functions. Instead of operators in the expansion of the fields, we
deal with $c$-number functions. These functions evolve in time in
such a way to satisfy the given initial fields distribution. We
have demonstrated that for ultrarelativistic neutrinos this method
also gives the appropriate description of spin-flavor oscillations
of Majorana neutrinos in external fields.

\section*{Acknowledgments}

I am grateful to the organizers of the $15^\mathrm{th}$
International Baksan School ``Particles and Cosmology -- 2011" for
the invitation and the financial support as well as to FAPESP
(Brazil) for a grant.

\end{document}